# Kinetics Modeling of Nanoparticle Growth on and Evaporation off Nanotubes

Vladimir Privman,[a] Vyacheslav Gorshkov,[b] and Yuval E. Yaish[c,*]

[a] Department of Physics, Clarkson University, Potsdam, NY 13699, USA
[b] National Technical University of Ukraine – KPI, Kiev 03056, Ukraine
[c] Andrew and Erna Viterbi Faculty of Electrical Engineering, Technion – Israel Institute of Technology, Haifa 3200003, Israel
* yuvaly@ee.technion.ac.il

**ABSTRACT**

A kinetic Monte Carlo approach is developed for studying growth and evaporation of nanoparticles on/off nanotubes. This study has been motivated by recent experimental advances in using nanoparticle evaporation (sublimation) off nanoparticle-decorated nanotubes for nanoscale "thermometry." We demonstrate that the considered kinetic Monte Carlo approach can reproduce features of the process that are not included in phenomenological thermodynamic modeling, as well as provide snapshots of the growth and evaporation process morphology.

# 1. INTRODUCTION

Carbon nanotubes (CNTs) and other low-dimensional carbon materials are promising components for future electronic devices.[1-4] Metallic CNTs can serve as interconnect electrodes for VLSI technology;[5] semiconducting CNTs may be utilized as the active channels of field effect transistors.[6,7] Understanding thermal properties of CNTs is essential for their applications in such platforms, and many studies have addressed this subject.[8] High electrical current in a suspended or on-substrate CNT introduces phonons and thermal effects, analysis of which is important in understanding transport properties.[9-13] Therefore, direct measurement techniques of the temperature variation profiles and time dependence are being presently actively developed.

Here we focus on the recently advanced novel approach involving utilization of nanoparticles (NP) grown[14] on CNTs as "local thermometers" at the nanoscale, with the thermal properties being probed by following the NPs' sublimation (evaporation) properties as a function of the temperature variation protocol. Modeling NP growth on and sublimation off CNTs can offer guidance for developing protocols to successfully monitor the temperature profile in order to enable accurate extraction of thermal parameters of CNTs, and to design optimal interfaces for thermal management, enabling applications in novel low-power electronics designs.[7]

Specifically, the design of the *growth process* can be optimized by kinetic Monte Carlo (MC) modeling predicting the time dependence of the particle sizes, particle-particle distance distribution statistics, and other properties dependent on the flux of matter and other controllable growth-process parameters, including the temperature. For the *evaporation stage*, process visualization and particle properties'—such as their linear dimensions—measurements can then be modeled. Here, the first demonstration-of-principle results of model calculations are reported.

It is important to emphasize that physical and transport properties depend on thermal "management" of the nanotubes. For example, at high bias, single wall CNTs exhibit current saturation,[9] and even negative differential resistance for suspended nanotubes.[10-12] This was attributed[10-12,15,16] to Joule heating, to electron-phonon scattering process due to zone boundary, and to nonequilibrium distribution of optical phonons.[17] When the current further increases, nanotube breakdown occurs.[18] This burning strategy was implemented[18] in order to eliminate metallic CNTs from semiconducting CNTs at different conductive junctions, and later was used[19] as a temperature variation signature for thermal analysis. Generally, indirect approaches to extracting thermal properties from transport data have been devised,[19,20] and data were collected to supplement such analysis.[21-28]

One can *directly measure the temperature* profile along the nanotube. This task is challenging due to small dimensions of CNTs. Promising approaches have been reported.[14,29,30] One technique involves evaporation of a small nanoparticle off the CNT under TEM imaging.[29] This procedure is suitable for on-substrate CNTs only, with a very thin TEM grid, and it is difficult to implement with a gating electrode. Another approach utilizes thermal scanning microscopy.[30] This method uses contact mode microscopy, and it was developed for on surface CNTs alone.[31] It is primarily sensitive to the lattice temperature, and less so to the optical-phonon temperature. In order to minimize nanotube damage while scanning, a thin film of insulating material covering the nanotube had to be utilized.[30]

A promising recent approach[14] that has motivated our study involves decoration of CNTs with nanoparticles, followed by optical imaging of the NPs' evaporation (sublimation). Growth or adsorption of nanoparticles on CNTs is generally well established.[32] Our aim has been to have a modeling approach for systems that have been experimentally identified as efficient and reproducible "thermometers" for probing thermal effects. Specifically, one can design[14] organic nanoparticles that are optically visible on suspended or on-substrate CNTs. Such NPs directly sublime from the solid to gas phase upon heating. Unlike small metallic NPs that were devised to act as binary temperature sensors,[29] the organic NPs are much larger, ~ 50–100 nm, and furthermore their sublimation rate depends on the temperature, and the sublimation process can be followed optically by dark-field microscopy.[14]

The outline of this work is as follows. In Sec. 2, we describe the kinetic MC modeling approach. Section 3 addresses a phenomenological thermodynamic model. Our results are presented in Sec. 4. Section 5 offers a short concluding statement.

## 2. KINETIC MONTE CARLO MODELING APPROACH

The kinetic MC approach utilized here, has originally been devised[33,34] to address the need for predicting the shape selection in the formation of even-proportioned—so called "isomeric"—crystalline NPs useful in catalysis and other applications.[35] It was also successfully applied to surface-templated growth of nanosize structures[36,37] and, recently, to processes of nanoparticle sintering[38,39] that involve not only growth but also *local dissolution* as particles by exchange of matter as they sinter into a single structure. Interestingly, studies of sintering have also involved[38] consideration of the effects of *temperature control protocols* on the process.

Here we outline the main features of this approach. NP growth and also dissolution in the studied systems have usually involved processes that are fast, i.e., nonequilibrium. The building-block atoms or molecules, to be termed "atoms" for brevity, are assumed to diffuse in the "gas" phase, and they can attach to the particle or detach from it. They can also locally hop on nanoparticle surfaces. One important assumption that makes such modeling numerically tractable involves the property that attachment of atoms is "registered" with the underlying lattice structure; for simplicity let us reference a monoatomic structure with coordination number $m_c$ and lattice spacing $a$. We address this assumption in the next paragraph. The "gas" atoms diffuse off-lattice by hopping at random angles in steps that are a fraction of *a*. Atoms already in a particle can move to nearby vacant sites or detach.

Atoms from the diffuser gas can attach at vacant lattice sites that are nearest-neighbor to the particle's atoms. This occurs once they hop into a Wigner-Seitz cell at such locations. Maintaining the precise "registration" of the attached atoms with the crystal lattice ensures[33,34,36-38] that we are considering nanocrystal morphologies of relevance for particles synthesized by fast nonequilibrium techniques.[35] Such particles do not have structure-spanning defects that can control shape variation by favoring the growth of certain crystalline faces and resulting in uneven-proportion shapes. Nanocrystal shapes of relevance here, cf. experimental work,[35,40] are approximately even-proportioned, "isomeric." In reality, large defects are avoided/not nucleated at the microscopic dynamics scales. However, for mesoscopic modeling the "exact registration" ensures the same result.[33,34,36-38] The shapes of "isomeric" nanocrystals thus grown are then bound[33,34] by lattice planes of symmetries similar to those in the equilibrium Wulff constructions,[41-43] but with different proportions. Monoatomic particle shapes for all the standard lattice symmetries were thus studied,[33,34] and the results were consistent with the experimental shapes of relevant metal/oxide nanocrystals[44] and core-shell noble-metal nanoparticles.[45,46]

In the present system particle growth is templated by the nanotube surface, and the "diffuser gas" is provided by CVD during the growth stage. Once the particles along the nanowire grow to the final sizes of interest, the CVD process is stopped. The resulting NP configuration, thus synthesized, is then evaporated at elevated temperatures.

Let us address specifics of the numerical MC modeling approach. Each attached crystal-lattice atom that is not fully blocked, can hop to vacant nearest neighbors (in our case; generally also to next nearest neighbors, etc., depending on the model choices[33]). The probabilities for such moves are proportional to temperature-dependent Boltzmann factors. Each unit MC time step constitutes a sweep through the system whereby all the diffuser-gas (detached) atoms are moved once on average, and attached lattice atoms have on average one hopping attempt. Hoppable atoms have coordination numbers $m_0 = 1, ..., m_c - 1$. The probability for them to move (i.e., for a hopping to be attempted) is $p^{m_0}$, corresponding to an activation free-energy barrier, $m_0 \Delta > 0$, were $p \propto e^{-\Delta/kT} < 1$. If the move is actually carried out, means, the atom went over this barrier, it will be repositioned to one of its $m_c - m_0$ vacant nearest neighbor sites or put back into it original site, with the probability proportional to the inverse of a free-energy change Boltzmann factor, i.e., to $e^{m_t|\varepsilon|/kT}$ (normalized over all the $m_c - m_0 + 1$ targets). Here $\varepsilon < 0$ is the free-energy measuring binding at the target sites. The

possible target site coordination can be $m_t = 1, \ldots, m_c - 1$ for hopping, and $m_t = 0$ for detachment. In the latter case, newly detached atoms rejoin the free diffuser gas.

Thus, in each unit-step sweep a sufficient number of random "probes" of atoms is carried out so that on average each atom is "probed" once per sweep, and moves are performed according to the aforementioned probabilistic rules:

- If the atom is in the gas, a diffusion step is carried out, and if the final position is within a unit cell near a crystal, the atom becomes crystal-lattice registered: part of a nanoparticle.
- If the atom is in the crystal and can hop (is not blocked all around), it is moved with probability determined by an activation free-energy barrier. The target site for the move is selected among the unblocked destination locations according to free-energy change Boltzmann factors. These destinations can be either nearest neighbor lattice sites that are part of the crystal (the atoms ends up connected) or a lattice-vector move can bring the atom to a position that is not connected to the crystal.
- In the latter case the atom is reclassified as an atom of the gas and freely diffuses next time it is "probed."
- The simulation is accelerated in various ways, for instance, by keeping and updating several lists of atoms with their and their neighbors' properties and their possible moves. The most obvious of these is the list of fully blocked crystal atoms, which cannot be moved and therefore can be ignored, with the appropriate reduction of the count of "probes per sweep," etc.

Additional information on the modeling methodology and specifics of setting up the kinetic MC approach are available, for instance, in Ref. 33, whereas details of how a substrate can "template" the growth process are summarized in Refs. 35, 38, with Refs. 33, 37 addressing setting up "boundary conditions" for supply of atoms. In our case, nanoparticles can only be initiated on (templated by) the line that models a nanotube: see Sec. 4. Initially all the atoms are in the gas for growth, in nanoparticles for dissolution: see Sec. 4.

Physically, we expect that the connected atoms' mobility is related to the surface diffusion coefficient which is proportional to $p$, set by the activation free-energy scale $\Delta$, such that

$$p \propto e^{-\Delta/kT}. \tag{1}$$

Another free-energy scale, $\varepsilon$, reflects local binding, and we use its magnitude scaled per $kT$ as follows,

$$\alpha = |\varepsilon|/kT. \tag{2}$$

Results[31,32,34-36] on particle synthesis and surface properties, etc., suggest that within the present setting typical nonequilibrium particle morphologies are maintained for a range of mesoscopic particle sizes if we assume reference values $\alpha_0$ and $p_0$ of order 1. Then the temperature can be increased (or decreased) by varying $\alpha$, with $p$ appropriately adjusted according to

$$p = (p_0)^{\alpha/\alpha_0}. \tag{3}$$

For evaporation, the temperature should be elevated as compared to synthesis, as was done for part of the duration of the particle-sintering process modeled by this approach.[36,37] Additional discussion of the modeling approach and its utility, as well as on the selection of parameter values is given in Sec. 4 and 5.

## 3. PHENOMENOLOGICAL THERMODYNAMIC MODEL FOR SUBLIMATION

The processes of NPs' synthesis, as they are grown on the CNTs, and their later evaporation (sublimation) are not in equilibrium for the particle as a whole (the processes are too fast for that), even though we just argued in Sec. 2 that, more locally, for specific atom's moves we can use energy-scale Boltzmann factors. However, it is instructive to consider simple equilibrium approaches,[47] supplemented with additional assumptions on the particle geometry, which yield useful phenomenological results.

Let us focus on the sublimation process that is used to monitor the temperature profile. Its rate depends on the partial pressure of the NP molecules in the gas phase in the vicinity of the NP.[47] Assuming a spherical (or an on-surface hemispherical) "drop" model for the NP, with radius $r$, one can employ the Kelvin equation for evaluating the partial pressure, $P_r$, in the gas phase adjacent to the NP surface:

$$P_r = P_\infty \exp\left(\frac{2\gamma_r M}{RT\rho r}\right), \tag{4}$$

where $P_\infty$ is the partial pressure above a flat surface, $\gamma_r$ is the surface energy (per unit area) of the NP, $\rho$ is its mass density, $M$ is its molar mass, and $R$ is the gas constant.

Defining $n_v$ as the number of molecules leaving a unit area of the NP per second, and $V_a$ as the volume per single molecule, we obtain the relation between the temporal change of the NP radius and the outgoing particle flux: $dr/dt = n_v V_a$. Kinetic theory suggests that the number of gas molecules colliding with the NP surface per unit area per second is $n_c = n\bar{c}/4$, where $n$ is the number of molecules per unit volume in the gas, and $\bar{c} = \sqrt{8RT/\pi M}$ is their mean speed. If we assume that the rates of condensation and evaporation are independent, then in dynamic equilibrium, $n_v = n_c f$, where $f$ is the fraction of colliding particles which become an integral part of the solid phase. Assuming that the dilute gas is *locally equilibrated near the surface*, and thus, we can use $n = P_r/kT$, the relations presented earlier in this paragraph combine to yield

$$\left.\frac{dr}{dt}\right|_r = \frac{P_r V_a f}{4kT}\sqrt{\frac{8RT}{\pi M}} = \frac{P_r f}{2\rho}\sqrt{\frac{2M}{\pi RT}} \propto T^{-1/2} P_r, \tag{5}$$

where in the last expression all the pre-factors (not shown) can be considered constant.

If the temperature, $T$, is also assumed kept constant on the time scales of the sublimation process, then, divided by the same relation in the limit of large radius, Eq. (5) yields a differential equation for $r(t)$ that is convenient to utilize in order to phenomenologically describe the NP sublimation process,

$$\left.\frac{dr}{dt}\right|_r = \left.\frac{dr}{dt}\right|_\infty \frac{P_r}{P_\infty} = \left.\frac{dr}{dt}\right|_\infty \exp\left(\frac{2\gamma_r M}{RT\rho r}\right). \tag{6}$$

The surface energy (per unit area) of a NP is generally assumed[48] somewhat lower than its bulk value, and depends on its size. This effect is usually represented via a Tolman-like relation

- 8 -

$$\gamma_r = \gamma_\infty \left(1 - \frac{2\delta}{r}\right), \tag{7}$$

where $\gamma_\infty$ is the bulk value, and the Tolman length, here $\delta > 0$, is comparable with the unit cell size of the sublimating material. When combined with Eq. (6), this yields a convenient expression for the sublimation rate,

$$\left.\frac{dr}{dt}\right|_r = -A\exp\left[B\left(1 - \frac{2\delta}{r}\right)\frac{1}{r}\right], \tag{8}$$

where $A = \left|\left(\frac{dr}{dt}\right)_\infty\right|$, and $B \simeq 2M\gamma_\infty/\rho RT$. Note that this equation is only meaningful for values of $r$ well over $\delta$. Thus, the linear size of the sublimating particle, modelled here as a "drop" of initial radius $r_0$, can be described by using the three-parameter phenomenological relation

$$\int_{r_0}^{r(t)} dr \exp\left(-B\left(1 - \frac{2\delta}{r}\right)\frac{1}{r}\right) = -At, \tag{9}$$

for $r_0 > r(t) \gg \delta$. In fact, a good estimate of the time it takes a particle to dissolve is obtained at $r(t) = 4\delta$, where $r = 4\delta$ is the inflection point of the $r$-dependence of the exponential in Eq. (8).

## 4. ILLUSTRATIVE MODELING RESULTS

We note that both models considered, the MC approach (Sec. 2) and the phenomenological description (Sec. 3) can in principle be viewed as involving various adjustable parameters. These are explicit in Eq. (9) for the phenomenological model, whereas for the MC model they are the two energy scales, $\Delta$ and $\varepsilon$, encountered in Eqs. (1), (2), the process time scale, as well as the specifics of lattice hopping rules and some other geometrical properties of the system incorporated, in the simulation. The data within the experimental approach considered here (illustrated below) are presently only preliminary,[14] whereas MC modeling is resource consuming. Therefore, in this work we limited ourselves to illustrative results that demonstrate that the MC approach can qualitatively reproduce general features of the data, including those that are not obtainable by the phenomenological data fitting with Eq. (9).

The latter fitting is illustrated for typical data, shown in Fig. 1, for sublimation of NPs of p-nitrobenzoic acid (pNBA). The phenomenological-model fit depicted in Fig. 1a does not reproduce a prominent feature in the data at small times, the origin of which will be addressed later. The morphology of this system as a function of time during sublimation, to the extent it can be resolved with AFM and optically, is illustrated in Fig. 2.

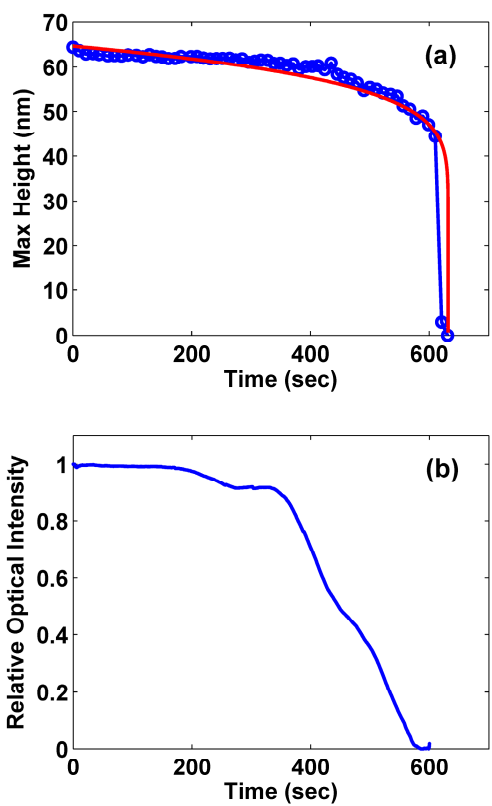

**Figure 1.** (a) The maximum NPs' heights measured by AFM as a function of time (blue circles) and its theoretical fit according to Eq. (9) (red line) at room temperature (295 K). (b) The relative dark field optical microscope intensity of the NPs on CNT as a function of time.

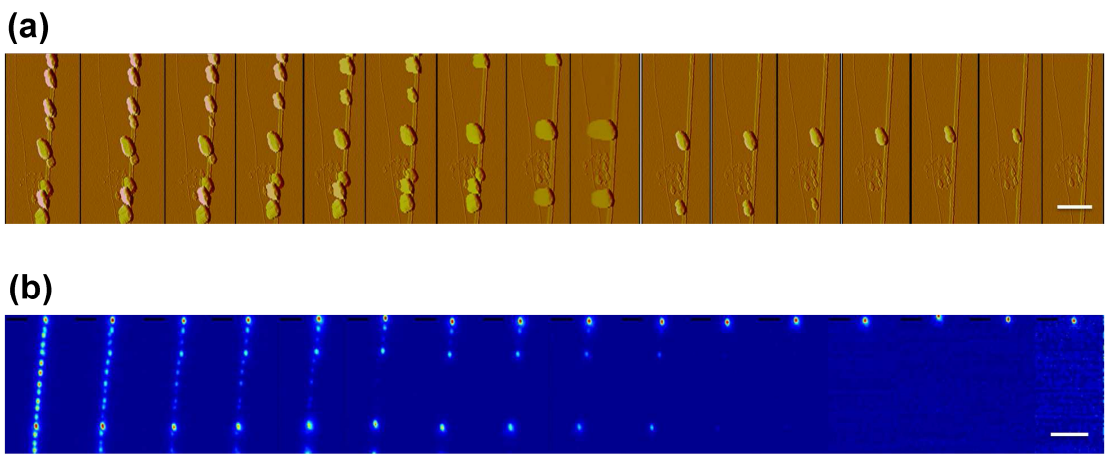

**Figure 2.** Nanoparticle sublimation: (a) A temporal sequence (left to right) of AFM amplitude images of pNBA NPs along a single CNT. The time interval between the images is 11 min. Scale bar is 1 µm. (b) A temporal sequence of dark field optical microscope NPs imaging intensity along a single CNT. The time interval between the images is 6.5 min. Scale bar is 4 µm.

Our kinetic MC modeling was carried out to accomplish a demonstration-of-principle that growth and evaporation of nanoparticles can indeed be studied with available numerical capabilities, and that features of evaporation process can be generally reproduced, including behavior not captured by the phenomenological thermodynamic model. We used our existing modeling codes (originally developed for metal nanoparticle synthesis and sintering) and a simplified model with the growth templated by the nanotube replaced by a line of surface atoms that do the templating (the initial attachment of atoms from the gas). Monoatomic FCC-lattice nanocrystals were then grown for convenient parameter choices in Eqs. (1), (2), with values discussed shortly, selected from the known[33,34,36-39] ranges. These parameters then yield isomeric nanocrystal growth and were qualitatively adjusted to mimic the experimentally observed particle proportions and spacing.

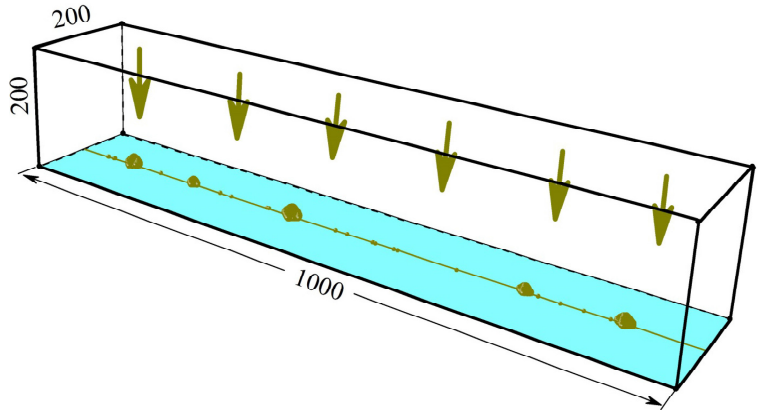

**Figure 3.** The geometry of the MC simulation; see text for details.

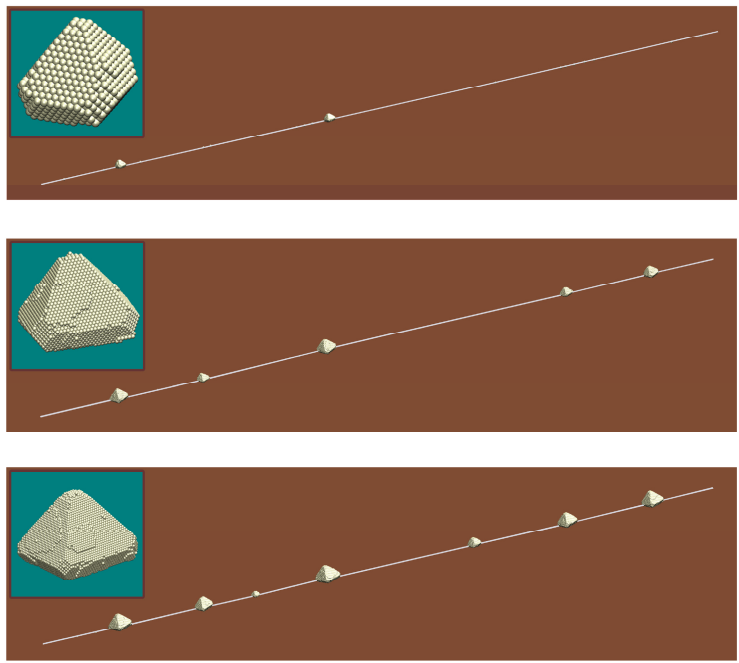

**Figure 4.** Nanocrystal growth modeled by kinetic MC, templated by a line on a surface, for MC-sweep time steps shown at times $t = 27 \times 10^5$ (top panel), $t = 50 \times 10^5$ (middle panel), and $t = 70 \times 10^5$ (bottom panel). The insets show the *leftmost* nanocrystal (at different magnifications), which for increasing times shown contains 2273, 14762, and 29972 atoms, respectively.

The geometry of a typical MC simulation is illustrated in Fig. 3. Similar to simulation of growth templated by a surface,[36,37] the atoms in the dilute gas were supplied during growth (schematically marked by arrow in Fig. 3) and removed during evaporation by keeping their concentration, $n_0$, constantly adjusted to a fixed value in a thin region (here 4 lattice layers) at the top of the box: $n_0 > 0$ for growth, $n_0 = 0$ for evaporation. Atoms in the gas diffuse be random-direction hopping (mentioned in Sec. 2) in the rest of the simulation box, here in steps of $1/\sqrt{2}$, where all the lengths (incl. in Fig. 3) are measured in terms of the FCC cubic-cell lattice spacing. All the other (but the top) box faces were reflecting for the atoms in the gas. However, a single middle lattice-line at the bottom of the box (see Fig. 3) is not reflecting, but rather atoms can attach at the lattice sites at that line and initiate particle growth.

Thus, nanocrystal growth is initiated on a line (Fig. 3). Results for growth with parameters $n_0 = 1.0 \times 10^{-3}$, $\alpha_0 = 1.5$, $p_0 = 0.6$, are shown in Fig. 4. The values for $\alpha_0$ and $p_0$ were selected from ranges identified in earlier studies[33,36-39] that lead to isomeric particle growth. These values were identified[33] for the present lattice choice (FCC) and hopping rule (only to nearest neighbors). The ranges were selected[33,36-39] to have isomeric particles (to avoid unstable, dendritic/fractal growth), with well-defined crystalline faces in proportions corresponding to those experimentally observed[44,49] in the non-equilibrium growth regime driven by supply of atoms. For this particular model, such growth can in principle be obtained with $p_0$ increased up to 0.9, as long as $\alpha_0$ is kept in the range $\alpha_0 \approx 0.7 - 2.5$, with specific value combinations also limited by the supply of matter, i.e., the choice of $n_0$, as well as by how the growth is initiated/templated. The set of values for Fig. 4 was selected based on preliminary studies aimed at qualitatively mimicking the few presently available experimental configurations (Fig. 2; Ref. 14). Later in this section, we illustrate what happens when $n_0$ is increased.

For times larger than $70 \times 10^5$ (the last image in Fig. 4) no new nanocrystals were nucleated, but those already formed further grew. Note that no single *gas atoms* are shown in Fig 4, but single atoms attached at the line are present here and in the later figures, though they are difficult to discern. However, Fig. 3 actually depicts the same growth stage as in middle panel of Fig 4, with the attached (but not gas) single atoms clearly visible, which was graphically done by using exaggerated-size (as compared to the lattice spacing) balls to represent all the shown atoms, which then caused some blurring of the larger clusters (in Fig. 3).

The growth process was then continued up to MC time $t = 103 \times 10^5$. This is the top panel in Fig. 5. The time count was then reset to $t = 0$, and evaporation was initiated by increasing the temperature: reducing the parameter $\alpha$ by 40%, and adjusting $p$ as per Eq. (3); removing the supply of the external matter that was replenishing the gas phase: $n_0 = 0$; and also, in fact, initially fully evacuating the gas from within the box at this point in time.

The resulting evaporation process is shown in Fig. 5 and in a movie accompanying it, the latter available as Supplemental Material, as well as in Fig. 6. Figure 5 depicts the actual particle images during the evaporation, whereas Fig. 6 shows how a particle's height decreases with time. Qualitatively, the modeling results are quite similar to those observed experimentally, including the sharp initial drop in the height (compare Fig. 6 to the data shown as blue circles in Fig. 1a). The latter feature will be revisited shortly.

Numerical simulations can be used to qualitatively explore parameter regimes of the problem and obtain interesting physical insights and pose interesting questions for experimental verification. For instance, if we increase the supply of atoms by 50%, to $n_0 = 1.50 \times 10^{-3}$, keeping other system parameters the same as before, we notice that the NPs are not only grown in a denser configuration, but are actually more uniform in their sizes, as shown in the top panel of Fig. 7. The latter observation would be interesting to confirm experimentally.

We also comment that the sharp initial drop in the particle height is present in this case as well, as seen in the bottom panel of Fig. 7. We note that this feature is not reproduced in the phenomenological model (Sec. 3). This is largely attributable to that the phenomenological model assumes a "drop" shape, whereas our results clearly suggest that during sublimation the nanoparticles are not only decreasing in size but also rounding up away from their original crystalline or polycrystalline shapes that were initially bounded by surface fragments close to crystalline-plane faces; Figs. 5, 6, 7. A more subtle additional effect is suggested by inspection of the distribution of the gas of "free atoms" near the line of the nanoclusters as the latter evaporate, while the atoms diffuse away from the nanoparticles, during the sublimation process. Indeed, for dense configurations (crowded up particles) excess concentration of the gas near the particles, amplified by the initially fast evaporation of sharp structural features can decrease the concentration gradient at the clusters' surfaces, which slows down evaporation for a while. As a result, the "shoulder" that follows the initial fast drop in the particle height can tend to flatten, as seen in Fig. 7 as compared to Fig. 6, as this in turn further emphasizes the initial-drop feature. Generally, in the present system, during sublimation the "gas" of atoms originates from the nanoparticles' own dissolution and is, in fact, to a certain degree centered at particles and densest at larger particles and particle clusters. This is difficult to visualize for parameter values that yield the process depicted in Fig. 7, because the gas is too dense. For the original process parameters, Figs. 5, 6, this behavior is illustrated in Fig. 8. The phenomenological model will have to be modified from its simplest variant considered in Sec. 3, at the expense of introducing additional processes and parameters, to accommodate such effects.

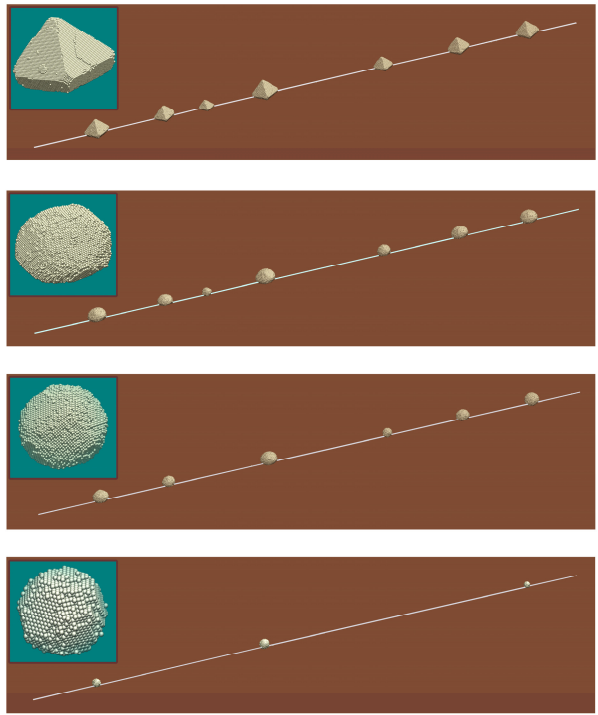

**Figure 5.** Nanocrystal evaporation after growth to time $t = 103 \times 10^5$ (top panel here; not shown in Fig. 4). Top panel: The evaporation was started (see text) and this time instance

was redefined as $t = 0$. Second panel: evaporation time $t = 4 \times 10^5$ MC-sweep time steps. Third panel: $t = 9 \times 10^5$. Bottom panel: $t = 19 \times 10^5$. The insets show the *leftmost* nanocrystal (at different magnifications), which for increasing times shown contains 63794, 47590, 32346, and 5877 atoms, respectively. These images highlight the fact that during evaporation the particles not only decrease in size, but their shapes round up. A movie of the evaporation process, the snapshots of which are presented here, is available as Supplemental Material. Note that a different cluster (the one that dissolves last) is magnified in the movie and single attached atoms are also shown.

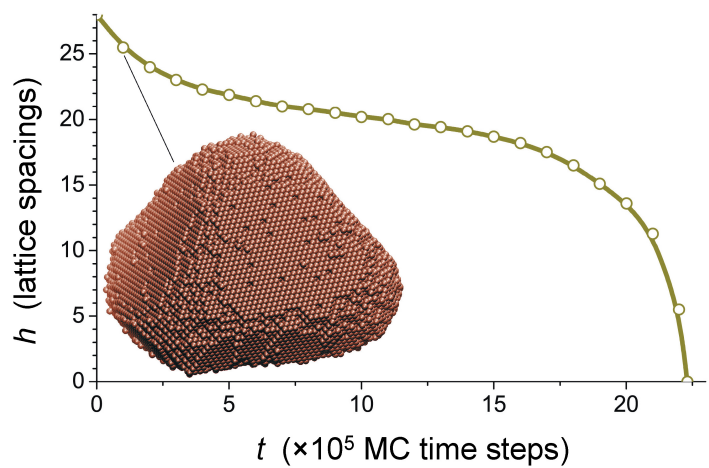

**Figure 6.** Variation of the height, $h$ (shown in terms of the FCC cubic-cell lattice spacing), of the *leftmost* cluster in the panels in Fig. 5, with time, until its compete evaporation. The inset here shows this cluster at a relatively short evaporation time, $t = 1 \times 10^5$, when its shape's rounding due to evaporation only begins to set in.

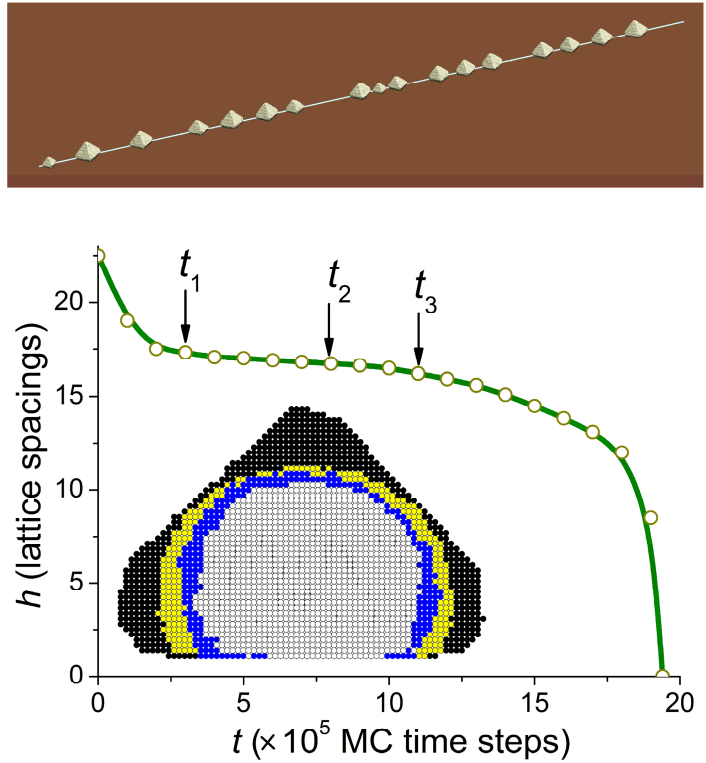

**Figure 7.** Top panel: Growth with the supply of atoms increased by 50% (see text), shown here at time $t = 52 \times 10^5$. Bottom panel: Evaporation of the fifth from the left nanoparticle in the configuration shown in the top panel. This nanoparticle was selected because it is rather large (has $3.42 \times 10^4$ atoms) and also positioned in a dense configuration (has relatively close neighbors), making it particularly suitable for illustrating the observation regarding the accumulating gas causing flattening of the shoulder after the initial sharp drop (see text). Its height is plotted as a function of time, and its cross-section in the plane that is along the "nanotube" line and perpendicular to the underlying surface, is shown in the inset, color-coded black, yellow, blue, gray for times $t = 0, t_1, t_2, t_3$ ($t_{1,2,3}$ are marked by arrows in the plot), respectively.

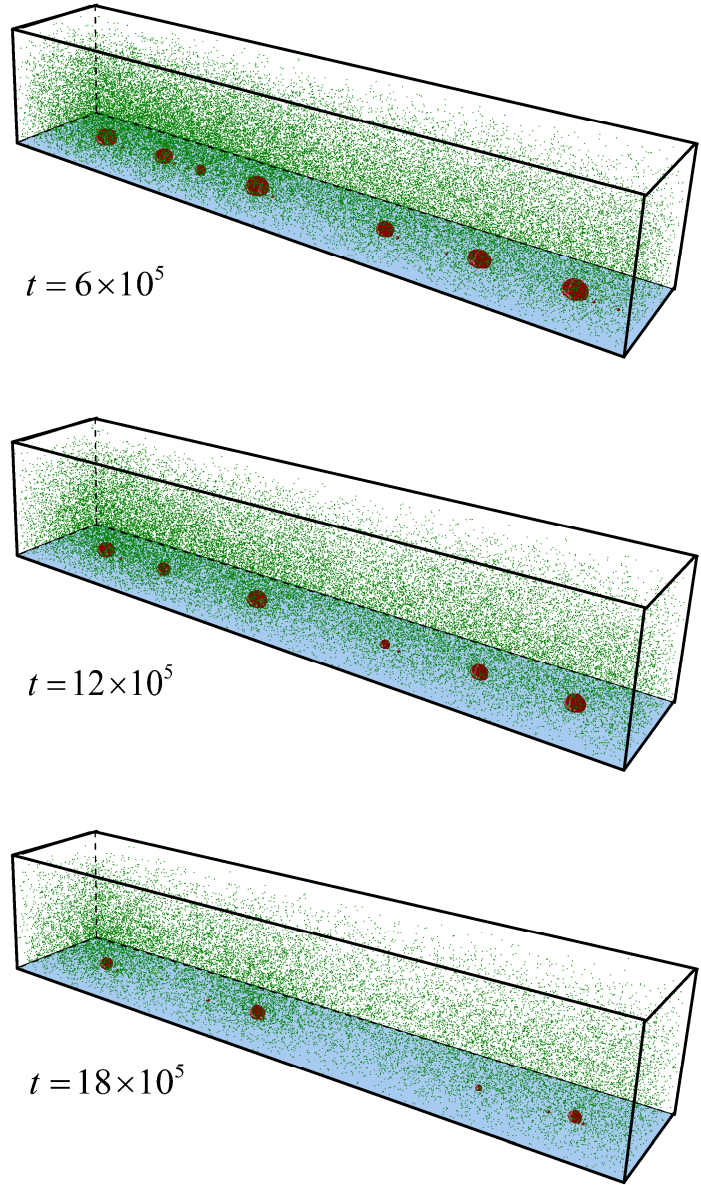

**Figure 8.** The sublimation process depicted in Figs. 5 and 6, is shown here for snapshots at three process times (different from those in Fig. 5). With proper color coding and selection of the image orientations, this figure visualizes the property that, the gas of atoms during sublimation, originating from the nanoparticles (recall that atoms reaching the top of the simulation box are removed, $n_0 = 0$), is to a certain extent centered at nanoparticles and is denser at larger particles and particle clusters.

## 5. CONCLUSION

We emphasize that at the present stage of our study the numerical kinetic MC results are qualitative, because we have to develop computer codes for more realistic substrates that model CNTs as the templates for growth, as well as correspond to the specific atoms or molecules of the NP materials, and crystal structures of the latter. For this, we need more definitive experimental results. Connecting the parameters of the present mesoscopic-scale model to more atomistic calculations, such as density functional theory, has not been attempted to our knowledge. In fact, no "atomistic" model of homogeneous nucleation, definitively explaining the emergence of the crystalline order has been accomplished to date. However, since the present system involves heterogeneous nucleation (templated by the nanotube), atomistic approaches might be successful, but such a multiscale study is outside the scope of the present article. Our results are promising in that they can reproduce features of the data, such as the short-time variation, and also offer process morphology visualization that is not possible with the "drop" phenomenological thermodynamic approach. In fact, kinetic MC modeling allows us to make not only sequences of snapshots of the processes involved, but also continuous movies of the time evolution.

## ACKMOWLEDGEMENTS

We wish to thank V. Kuzmenko for carrying out some of the programming and simulation work and G. Zeevi and S. Rechnitz for the contributing to measurements of the illustrative experimental data.

## SUPPLEMENTAL MATERIAL

See Supplemental Material (after the references in this e-print) for a movie of the evaporation process the snapshots of which are presented in Fig. 5.

**SUPPLEMENTAL MATERIAL:** Click the arrow to play the movie (will work in most PDF readers, but not from within browsers).

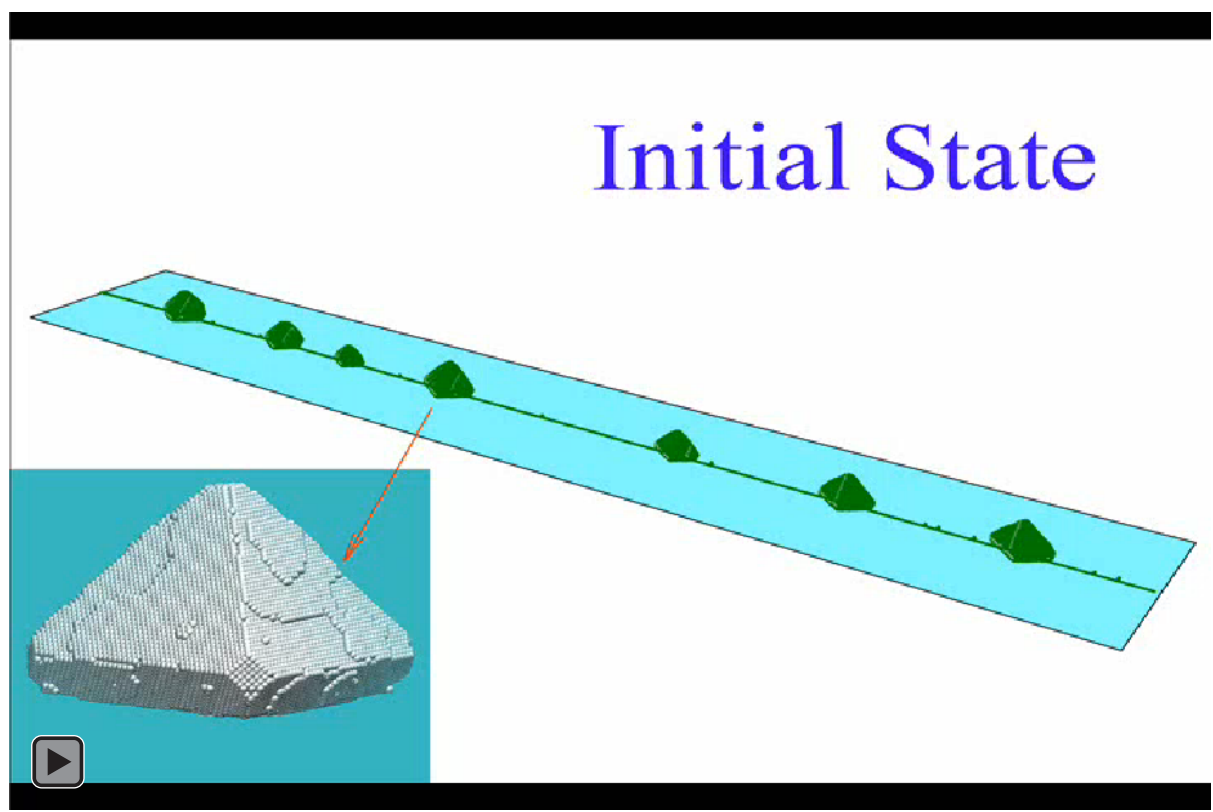